\documentclass[pra,twocolumn,floatfix,superscriptaddress,notitlepage,longbibliography]{revtex4-1}

\usepackage[normalem]{ulem}
\usepackage{graphicx, color, graphpap}
\usepackage{enumitem}
\usepackage{amssymb}
\usepackage{amsthm}
\usepackage{multirow}
\usepackage[colorlinks=true,citecolor=blue,linkcolor=magenta]{hyperref}
\usepackage[T1]{fontenc}
\usepackage{verbatim}
\usepackage{mathtools}
\usepackage{titlesec}

\long\def\ca#1\cb{} 

\newcommand{\abs}[2][]{#1| #2 #1|}
\newcommand{\braket}[2]{\langle #1 \hspace{1pt} | \hspace{1pt} #2 \rangle}

\newcommand{\bramatket}[3]{\langle #1 \hspace{1pt} | #2 | \hspace{1pt} #3 \rangle}

\newcommand{\norm}[2][]{#1| \! #1| #2 #1| \! #1|}

\newcommand{\ket}[1]{|#1\rangle}               
\newcommand{\bra}[1]{\langle #1|}              
\newcommand{\dya}[1]{\ket{#1}\!\bra{#1}}

\newcommand{\poly}{\operatorname{poly}}

\newcommand{\GHZ}{\text{GHZ}}

\newcommand{\GC}{\mathcal{G}}

\newcommand{\IC}{\mathcal{I}}

\newcommand{\LC}{\mathcal{L}}

\newcommand{\OC}{\mathcal{O}}

\newcommand{\Tr}{{\rm Tr}}

\newcommand{\Var}{{\rm Var}}

\renewcommand{\geq}{\geqslant}
\renewcommand{\leq}{\leqslant}

\newcommand{\Rbb}{\mathbb{R}}

\renewcommand{\vec}[1]{\boldsymbol{#1}}  

\newcommand{\ad}{^\dagger}

\newcommand*{\id}{\openone}

{}
{}

\newtheorem{definition}{Definition}

\begin{document}
\title{Variational Quantum Algorithm for Estimating the Quantum Fisher Information}

\author{Jacob L. Beckey}
\affiliation{Theoretical Division, Los Alamos National Laboratory, Los Alamos, New Mexico 87544}
\affiliation{JILA, NIST and University of Colorado, Boulder, Colorado 80309}
\affiliation{Department of Physics, University of Colorado, Boulder, Colorado 80309}

\author{M. Cerezo}
\affiliation{Theoretical Division, Los Alamos National Laboratory, Los Alamos, New Mexico 87544}
\affiliation{Center for Nonlinear Studies, Los Alamos National Laboratory, Los Alamos, New Mexico 87544}
\affiliation{Quantum Science Center, Oak Ridge, TN 37931, USA}

\author{Akira Sone}
\affiliation{Theoretical Division, Los Alamos National Laboratory, Los Alamos, New Mexico 87544}
\affiliation{Center for Nonlinear Studies, Los Alamos National Laboratory, Los Alamos, New Mexico 87544}
\affiliation{Quantum Science Center, Oak Ridge, TN 37931, USA}
\affiliation{Aliro Technologies, Inc. Boston, Massachusetts 02135, USA}

\author{Patrick J. Coles}
\affiliation{Theoretical Division, Los Alamos National Laboratory, Los Alamos, New Mexico 87544}
\affiliation{Quantum Science Center, Oak Ridge, TN 37931, USA}

\begin{abstract} 
 The Quantum Fisher information (QFI) quantifies the ultimate precision of estimating a  parameter from a quantum state, and can be regarded as a reliability measure of a quantum system as a quantum sensor. However, estimation of the QFI for a mixed state is in general a computationally  demanding task. In this work we present a variational quantum algorithm called Variational Quantum Fisher Information Estimation (VQFIE) to address this task. By estimating lower and upper bounds on the QFI, based on bounding the fidelity, VQFIE outputs a range in which the actual QFI lies. This result can then be used to variationally prepare the state that maximizes the QFI, for the application of quantum sensing. In contrast to previous approaches, VQFIE does not require knowledge of the explicit form of the sensor dynamics.  We simulate the algorithm for a magnetometry setup and demonstrate the tightening of our bounds as the state purity increases.  For this example, we compare our bounds to literature bounds and show that our bounds are tighter. 
\end{abstract}
\maketitle

\section{Introduction}
The goal of quantum sensing is to utilize quantum coherence or quantum entanglement to better estimate unknown parameters of quantum systems via measurement~\cite{Pezze09,giovannetti2011advances,Degen16x}. This includes quantum magnetometry~\cite{Taylor08, Dutta20}, quantum thermometry~\cite{Sone18a,Sone19a,correa2015individual,Pasquale16}, quantum illumination~\cite{Lloyd2008,Zhuang_2017}, distributed sensing~\cite{zhuang2020distributed,proctor2018multiparameter}, quantum system identification~\cite{Burgarth12} for the estimation of the Hamiltonian parameters~\cite{Zhang14, burgarth2009indirect, Sone17a, Franco09,Wang20}, graph structure~\cite{Kato14} or system dimensions~\cite{Sone17b, Owari15}. The methodologies developed in quantum sensing are expected to contribute to the progress in various state-of-the-art fields of science and technology, such as molecule structure determination~\cite{Ajoy17,Lovchinsky16},  biosensing~\cite{Choi20,Fujiwara20}, nanomaterial magnetism~\cite{Wolfe16,Casola18}, dark matter detection~\cite{Rajendran17}, and gravitational wave detection~\cite{McCuller2020,Abbott2016}.

Quantum Fisher Information (QFI) is a fundamentally important quantity in quantum sensing because it quantifies the ultimate precision achievable in estimating a parameter $\theta$ from a quantum state $\rho_{\theta}$ via the quantum Cram{\'{e}}r-Rao bound (QCRB)~\cite{Hayashi_2004, Jing20}. For single parameter estimation, QFI is associated with the standard fidelity between the true state $\rho_\theta$ and an error state, $\rho_{\theta+\delta}$. The intuition behind this relation is that QFI captures the response of the quantum state to a small change in~$\theta$. A true state with a high QFI will be very distinguishable from the error state, making it easier to estimate the parameter via measurement.

While the QFI has been extensively studied for pure quantum states \cite{fujiwara1995quantum,matsumoto2002new}, the case of mixed states has received considerably less attention, with recent theoretical results shedding light on using mixed states for metrology applications~\cite{fiderer2019maximal}. Since preparing pure quantum states is intrinsically a difficult task (due to system-environment interactions), most quantum states prepared in quantum hardware are mixed states. As such, one of the main goals for state-of-the-art technologies for quantum control is the reduction of quantum noise to be able to prepare states with high purities.      

An example of such technologies are quantum computers. While quantum devices are expected to outperform classical computers in many tasks such as factoring and simulating complex systems, currently available quantum devices, known as Noisy Intermediate-Scale Quantum (NISQ) computers~\cite{preskill2018quantum}, are prone to hardware noise and hence prepare mixed states. Moreover, the limited number of qubits and constrained circuit depth makes it impractical to implement error correction schemes. This does not preclude, however, the possibility of still employing mixed states (with either high purity or high rank) for practical applications, and this is precisely the scope of this work.

Variational Quantum Algorithms (VQAs) are one of the most promising strategies to overcome these limitations in the NISQ era \cite{cerezo2020VQAreview} and make use of near-term quantum devices. In VQAs, a cost function $C({\vec{\alpha}})$ is efficiently estimated with a quantum computer, while part of the computational complexity is pushed to a classical optimizer which minimizes the cost by adjusting the parameters $\vec{\alpha}$ of a parameterized quantum circuit. VQAs have been studied for various applications~\cite{VQE, farhi2014quantum, Romero, QAQC, VQSD, arrasmith2019variational, cerezo2019variational, sharma2019noise, bravo2020quantum, cerezo2020variational, cirstoiu2020variational, yuan2019theory,huang2020GANs,carolan2020variational, bravo-prieto2019, anschuetz2019variational}, and the scaling of their trainability has been explored recently~\cite{mcclean2018barren, cerezo2020cost,sharma2020trainability,volkoff2020large,wang2020noise,cerezo2020impact,holmes2020barren}. 

In the past decade, there have been several proposals to apply classical machine learning methods to the quantum parameter estimation problem \cite{hentschel2010machine,xu2019generalizable,xiao2019continuous}.  However, even more recently, the prospect of instead using NISQ devices to enhance quantum sensing capabilities has become an exciting research direction and is precisely the topic of our work. State preparation for sensing via VQAs has been proposed for phase estimation in trapped atomic arrays~\cite{Kaubruegger19,LucaReview}, noisy magnetometry~\cite{koczor2020variational}, multiple parameter estimation~\cite{Meyer20}, and for phase estimation assisted by purity loss measurement~\cite{Modi16,Yang20}. However, it is not obvious that these protocols avoid an important practical issue known as barren plateaus in the cost training landscape~\cite{mcclean2018barren,cerezo2020cost}. In addition, they can also require detailed information about the dynamics of how the parameter $\theta$ is encoded in the quantum system, which is not always known in practice, and which is not needed in our work.

\begin{figure}
\centering
\includegraphics[width=0.48\textwidth]{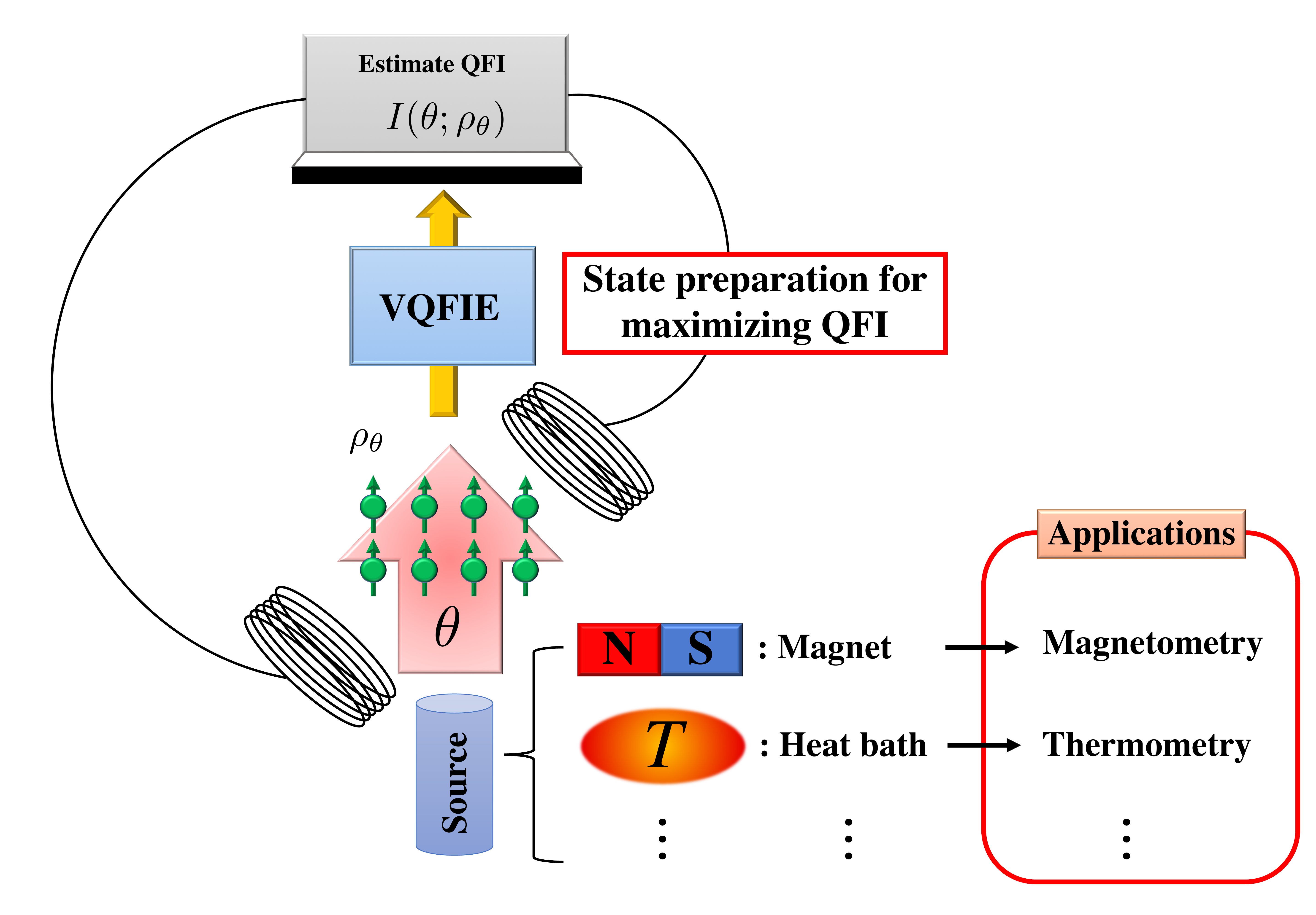}
\caption{\textbf{Application of the Variational Quantum Fisher Information Estimation (VQFIE) algorithm.}
A quantum system $\rho$ interacts with a source that encodes the information of a  parameter $\theta$ in the state as $\rho_{\theta}$. The goal of VQFIE is to estimate the quantum Fisher information (QFI), $I(\theta;\rho_{\theta})$, which is related to the minimal achievable uncertainty when  estimating $\theta$ from $\rho_\theta$. One can then use VQFIE to variationally prepare the state $\rho$ that maximizes the estimated QFI. The VQFIE algorithm can then be used in many applications related to precision sensing, such as magnetometry and thermometry. }
\label{fig:sensing}
\end{figure}

Here we propose a VQA to estimate the QFI on mixed states that addresses the issues previously mentioned. Namely, our method could avoid barren plateaus, and does not require information about the dynamics of interest (i.e. the explicit mathematical form of the generator of unitary dynamics is not required). We name this algorithm  the Variational Quantum Fisher Information Estimation (VQFIE) algorithm.  VQFIE computes upper and lower bounds on the QFI, and these bounds are based on bounding the quantum fidelity. We specifically focus on bounds obtained by truncating the spectrum of the exact state~\cite{cerezo2019variational}, which can be computed by taking advantage of previous variational methods for obtaining a quantum state's principal components~\cite{cerezo2020variational}. One can then use VQFIE to variationally prepare the  state that maximizes the estimated QFI. As schematically shown in Fig.~\ref{fig:sensing}, we expect applications of our proposed algorithm in various fields such as material science, biology, and chemistry.

The paper is organized as follows. In Sec.~\ref{sec:bounds_VQFIE}, we formulate the basic theory for VQFIE by introducing novel lower and upper bounds on the QFI. Then, in Sec.~\ref{sec:VQFIE} we present the structure of the VQFIE algorithm for computing the aforementioned bounds and for estimating the QFI. In Sec.~\ref{sec:numericalsimulation}, we present numerical simulations of VQFIE for a magnetometry application. We finally compare our bound to the literature in Sec.~\ref{sec:lowerboundcomparison}, followed by the concluding remarks in  Sec.~\ref{sec:discussion}. 

\section{Theoretical Framework}
\label{sec:bounds_VQFIE}

\subsection{General background}
\label{sec:background}

Consider the scenario where an $n$-qubit quantum state $\rho$, known as the probe state, interacts with a source that encodes the information of a single parameter $\theta$ and maps the input state into the so-called exact state  $\rho_\theta$. The QFI quantifies the ultimate precision $\Delta \theta$ when estimating the  parameter $\theta$ from  $\rho_{\theta}$ via the Quantum Cram\'er-Rao Bound as~\cite{Hayashi_2004, Jing20} 
\begin{align}\label{eq:precision}
(\Delta \theta)^2\geq \frac{1}{\nu I\left(\theta;\rho_{\theta}\right)}\,,
\end{align}
where $\nu$ is number of measurement repetitions used to estimate $\theta$, and where $I\left(\theta;\rho_{\theta}\right)$ is the QFI which is uniquely determined by the parameter to be estimated and the measured quantum state. We note that the standard mathematical definition of the QFI is given in terms of a so-called symmetric logarithmic derivative (SLD) operator, which need not be unique in general \cite{Liu16QFIvsSLD}. However, here we utilize a more experimentally useful definition of the QFI for single parameter estimation, given as ~\cite{Hayashi_2004, Jing20}
\begin{equation}
\begin{split}
    I(\theta;\rho_{\theta})    =-4\lim_{\delta\to 0}\partial_{\delta}^2 F(\rho_{\theta},\rho_{\theta+\delta})\,,
    \label{eq:fidelityQFI0}
\end{split}
\end{equation}
where $F(\rho,\sigma)=\norm{\sqrt{\rho_{\theta}}\sqrt{\rho_{\theta+\delta}}}_1=\Tr\left[\sqrt{\sqrt{\rho_{\theta}} \rho_{\theta+\delta} \sqrt{\rho_\theta}}\right]$ is the standard fidelity between the exact state $\rho_{\theta}$ and the error state  $\rho_{\theta+\delta}$.

Equation~\eqref{eq:fidelityQFI0} quantifies the sensitivity of the state $\rho$ to small changes $\delta$ in the parameter as the second partial derivative of the fidelity between exact and error states. Hence, the more sensitive $\rho_{\theta}$ is to these small $\theta$ changes, the larger the QFI is, and the more precise the estimation of the parameter will be according to Eq.~\eqref{eq:precision}. Note that here no assumptions were made regarding what $\theta$ is, or how it was encoded in $\rho_{\theta}$. This formalism then encompasses cases such as $\theta$ being the  magnitude of a field (magnetometry) or a temperature (thermometry). 

In practice one approximates the QFI by
\begin{align}
    I_\delta(\theta;\rho_{\theta})= 8\frac{1-F(\rho_{\theta},\rho_{\theta+\delta})}{\delta^2}\,, \label{eq:fidelityQFI}
\end{align}
where $|\delta|\ll 1$. Although, a smaller $\delta$ will always lead to a better approximation, the achievable range of values will depend on the experimental implementation. In the limit of the parameter shift $\delta$ approaching zero, Eq.~\eqref{eq:fidelityQFI} becomes the QFI as $I(\theta;\rho_{\theta})=\lim_{\delta\to 0}I_\delta(\theta,\rho_{\theta})$, because the fidelity takes its maximum at $\delta=0$~\cite{Hayashi_2004,Jing20}. For pure states such quantity can be efficiently computed on a quantum computer, as the fidelity between two pure states is simply given by their overlap, i.e., $F(\ket{\psi_\theta},\ket{\psi_{\theta+\delta}})=|\braket{\psi_\theta}{\psi_{\theta+\delta}}|$. However, for general mixed state there is no efficient algorithm to directly compute the standard fidelity and the QFI in~\eqref{eq:fidelityQFI}. This does not preclude the possibility of estimating the QFI by calculating efficiently computable upper and lower bounds of  $I_\delta(\theta;\rho_{\theta})$, which is precisely the goal of VQFIE.

\subsection{Bounds on the Quantum Fisher Information}
\label{sec:TQFIB}

There are many ways in which to bound and estimate the QFI \cite{GirolamiPRL,GirolamiExpPRA,Yang20,Modi16,LucaReview, katariya2021geometric} however many have scaling that precludes their implementation on near-term devices. Moreover, most known methods require detailed knowledge of how the unknown parameter was encoded in the state, which is certainly not always known in practice. For a VQA intended to run on NISQ hardware, quantities of interest must be computable with sufficiently shallow quantum circuits. In this section we present such bounds on the QFI which are derived from upper and lower bounding the standard fidelity in Eq.~\eqref{eq:fidelityQFI}. 

That is, for any two functions $f_1(\rho_{\theta},\rho_{\theta+\delta})$ and $f_2(\rho_{\theta},\rho_{\theta+\delta})$ such that
\begin{equation}\label{eq:generalfidbounds}
    f_1(\rho_{\theta},\rho_{\theta+\delta})\leq F(\rho_{\theta},\rho_{\theta+\delta}) \leq f_2(\rho_{\theta},\rho_{\theta+\delta})\,,
\end{equation}
we can obtain induced bounds for the QFI  as
\begin{equation}\label{eq:QFIbounds}
   \IC_\delta\left(f_2;\rho_{\theta}\right)\leq I_\delta\left(\theta;\rho_{\theta}\right) \leq \IC_\delta\left(f_1;\rho_\theta\right)\,,
\end{equation}
where we defined the induced bound for a function $f(\rho_{\theta},\rho_{\theta+\delta})$ as
\begin{equation}\label{eq:generalfunction}
  \IC_\delta\left(f; \rho_\theta\right)=   8\frac{1-f(\rho_{\theta},\rho_{\theta+\delta})}{\delta^2}\,,
\end{equation}
which in turn allows us to define bounds for the QFI in the $\delta\rightarrow 0$ limit as $\IC\left(f;\rho_{\theta}\right)=\lim_{\delta\to 0}\IC_\delta\left(f; \rho_\theta\right)$.

In what follows, we first summarize our recent result of a lower and upper bound for the QFI called the truncated QFI (TQFI) bounds~\cite{Sone2020QFI}, which are based on truncating the exact state $\rho_\theta$ to its largest $m$ eigenvalues~\cite{cerezo2019variational} and computing the truncated fidelities~\cite{cerezo2019variational}.
Then, we employ the so-called sub- and super-Fidelities~\cite{MiszczakPHUZ09} to derive the {\it Sub- and Super-Quantum Fisher Information} (SSQFI) bounds. 
Here we remark that, as discussed below, the lower TQFI bound is not just a bound on the QFI, but has additional operational meaning.

\subsubsection{Truncated QFI}
Let $\rho_{\theta}=\sum_{k=1}^r \lambda_k\dya{\lambda_k}$ be the spectral decomposition of the exact state, where   $\lambda_k$ is the $k$-th the eigenvalue of $\rho$, and $\ket{\lambda_k}$ its associated eigenvector. Here, $1\leq r\leq 2^n$ is the rank of $\rho_{\theta}$.  Moreover, let us assume that the eigenvalues are ordered in decreasing order such that $\lambda_k\geq \lambda_{k+1}$.
Then, for a given integer $m$ such that $1\leq m\leq r$, we define the sub-normalized states
\begin{align}\label{eq:truncated-state}
    \rho^{(m)}_{\theta}  &= \Pi^{m}_{\rho_{\theta}}\rho_{\theta}\Pi^{m}_{\rho_{\theta}}=\sum_{k=1}^{m}\lambda_k\dya{\lambda_k}\,,\\
    \rho^{(m)}_{\theta+\delta}  &= \Pi^{m}_{\rho_{\theta}}\rho_{\theta+\delta}\Pi^{m}_{\rho_{\theta}}\,,
\end{align}
where  $\Pi^{m}_{\rho_{\theta}}  = \sum_{k=1}^{m}\dya{\lambda_k}$. That is, $\rho^{(m)}_{\theta}$ and $\rho^{(m)}_{\theta+\delta}$ are respectively obtained by  projecting the exact and error states into the subspace generated by the $m$ eigenvectors of $\rho_{\theta}$ associated to its $m$ largest eigenvalues.

As shown in~\cite{Tomamichel_2016,cerezo2019variational}, the following bounds hold $\forall\delta\in\mathbb{R}$
\begin{align}
    F\left(\rho_{\theta}^{(m)},\rho_{\theta+\delta}^{(m)}\right) \leq F\left(\rho_{\theta},\rho_{\theta+\delta}\right)\leq F_{*}\left(\rho_{\theta}^{(m)},\rho_{\theta+\delta}^{(m)}\right)\,.
    \label{eq:TFB1}
\end{align}
Here $ F\left(\rho_{\theta}^{(m)},\rho_{\theta+\delta}^{(m)}\right)=\norm{\sqrt{\rho_{\theta}^{(m)}}\sqrt{\rho_{\theta+\delta}^{(m)}}}_1$ is the {\it truncated fidelity}, and $F_{*}\left(\sigma,\tau\right)$ denotes the {\it truncated generalized fidelity} between two sub-normalized states $\sigma$ and $\tau$, given by  
\begin{align}\label{eq:truncatedgeneralizedfidelity}
    F_{*}\left(\sigma,\tau\right)=&\norm{\sqrt{\sigma}\sqrt{\tau}}_1
    +\sqrt{\!\left(1\!-\!\Tr\left[\sigma\right]\right)\!\!\left(1\!-\!\Tr\left[\tau\right]\right)}\,.
\end{align}
Here we remark that the bounds in~\eqref{eq:TFB1} get monotonically tighter as $m$ increases, with equalities holding if $m=r$~\cite{cerezo2019variational}. We note that, as discussed below, the truncated fidelity bounds can be computed with $2n+1$ qubits. Additionally, as shown in~\cite{cerezo2019variational} the bounds in~\eqref{eq:TFB1} are tight if: (1) $\rho$ is a high purity state, or (2) $\rho$ is a low rank state and if $m=r$. 

Combining Eqs.~\eqref{eq:generalfidbounds}--\eqref{eq:generalfunction} with the truncated fidelity bounds in~\eqref{eq:TFB1} allows us to define the TQFI bounds~\cite{Sone2020QFI}  
\begin{equation} \label{eq:TQFI-bounds}
   \IC_\delta\left(F_{*}; \rho_{\theta}^{(m)}\right)\leq I_{\delta}(\theta;\rho_{\theta}) \leq \IC_\delta\left(F; \rho_\theta^{(m)}\right)\,,
\end{equation}
with equalities again holding if $m=r$. We note that the quantity $\IC(F_{*}; \rho_{\theta}^{(m)})$  was recently introduced in~\cite{Sone2020QFI} and is known as the  Truncated Quantum Fisher Information (TQFI). The TQFI represents a generalization of the QFI for sub-normalized states as it satisfies the canonical criteria of a QFI measure. We encourage the interested reader to see Ref.~\cite{Sone2020QFI} for details.

\subsubsection{Sub- and super-bounds}

As shown in~\cite{MiszczakPHUZ09}, the following bounds hold $\forall\delta\in\mathbb{R}$ 
\begin{align}\label{eq:SSFB}
    \sqrt{E(\rho_{\theta},\rho_{\theta+\delta})}
    \leq F(\rho_{\theta},\rho_{\theta+\delta})
    \leq \sqrt{R(\rho_{\theta},\rho_{\theta+\delta})}\,.
\end{align}
Here, $E(\rho,\sigma)$ and $R(\rho,\sigma)$ are respectively called the {\it sub-fidelity}, and {\it super-fidelity} between the quantum states $\rho$ and $\sigma$, and are defined as 
\begin{align}\label{eqn:subfid}
    E(\rho,\sigma)&=\Tr\left[\rho\sigma\right]+\sqrt{2\left(\left(\Tr\left[\rho\sigma\right]\right)^2-\Tr\left[\rho\sigma\rho\sigma\right]\right)}\,,\\
    \label{eqn:superfid}
    R(\rho,\sigma)&=\Tr\left[\rho\sigma\right]+\sqrt{\left(1-\Tr\left[\rho^2\right]\right)\left(1-\Tr\left[\sigma^2\right]\right)}\,.
\end{align}
As shown in Appendix \ref{sec:estimatingFunctionals}, because these quantities are expressed as traces of products of quantum states, they can be efficiently estimated on a quantum computer requiring up to $4n+1$ qubits~\cite{cerezo2019variational,MiszczakPHUZ09,ekert2002direct}.

By combining Eqs.~\eqref{eq:generalfidbounds}--\eqref{eq:generalfunction} with the sub-, and super-fidelity bounds in~\eqref{eq:SSFB} we can define the Sub- and Super-Quantum Fisher Information (SSQFI) bounds
\begin{equation}\label{eq:ssqfibounds}
   \IC_\delta\left(\sqrt{R};\rho_{\theta}^{(m)}\right)\leq I_\delta(\theta;\rho_{\theta}) \leq \IC_\delta\left(\sqrt{E}; \rho_\theta^{(m)}\right)\,.
\end{equation}

It is worth noting that it has been recently shown in~\cite{cerezo2021sub} that the sub-QFI bound is faithful to the QFI in the sense that both quantities are maximized and minimized for the same quantum states. Such a result implies that the state found by  maximizing the lower bound in~\eqref{eq:ssqfibounds} is optimal for metrology and sensing applications.

\subsubsection{Dynamics-agnostic QFI bounds}

One of the main advantages of the TQFI and SSQFI bounds is that that they are dynamics agnostic, meaning that their computation requires no knowledge about how the source encodes the parameter $\theta$ in  $\rho$. This is in contrast to other known bounds on the QFI. For instance, for phase estimation where the state is given by~\eqref{eq:phase} one can show that $4(\Tr[\rho_\theta^2 G^2]-\Tr[(\rho_{\theta} G)^2])$ is a lower bound, and $4(\Tr[\rho_\theta G^2]-(\Tr[\rho_{\theta} G])^2)$ is an upper bound on $I(\theta;\rho_{\theta})$~\cite{Modi16}. However, computing these bounds requires knowledge of the generator $G$, which might not always be accessible. To avoid requiring such extra knowledge, we here instead  define the following dynamics-agnostic quantities from the bounds  previously presented:
\begin{align}\label{eq:H_delta}
  H_{\delta}(\theta;\rho_{\theta}) &= \max\left\{ \IC_\delta\left(F_{*},\rho_\theta^{(m)}\right),\IC_\delta\left(\sqrt{R};\rho_{\theta}^{(m)}\right)\right\},  \\
  \label{eq:J_delta}
   J_{\delta}(\theta;\rho_{\theta}) &= \min\left\{\IC_\delta\left(F,\rho_\theta^{(m)}\right),\IC_\delta\left(\sqrt{E}; \rho_\theta^{(m)}\right)\right\}\,.
\end{align}
Hence, the following bounds on the QFI hold 
\begin{align}\label{eq:VQFIEbounds}
    H_{\delta}(\theta;\rho_{\theta})\leq  I_\delta (\theta;\rho_{\theta}) \leq J_{\delta}(\theta;\rho_{\theta})\,.
 \end{align}
 
It is  worth noting that as shown in~\cite{cerezo2019variational}, the TQFI bounds are often tighter than the SSQFI bounds when $m\in\OC(\poly(n))$. Hence, for large enough $m$ we will have $H_{\delta}(\theta;\rho_{\theta}) = \IC_\delta\left(F_{*},\rho_\theta^{(m)}\right)$ and $J_{\delta}(\theta;\rho_{\theta}) = \IC_\delta\left(F,\rho_\theta^{(m)}\right)$, meaning that it will suffice to compute the TQFI bounds. Moreover, as previously mentioned, the computation of the TQFI bounds requires only $2n+1$ qubits for $n$-qubit states $\rho$, while the computation of the SSQFI bounds (specifically the upper bound) requires $4n+1$ qubits.

\section{Variational Quantum Fisher Information Estimation (VQFIE) algorithm}
\label{sec:VQFIE}

\begin{figure}[t]
    \includegraphics[width=0.9\columnwidth]{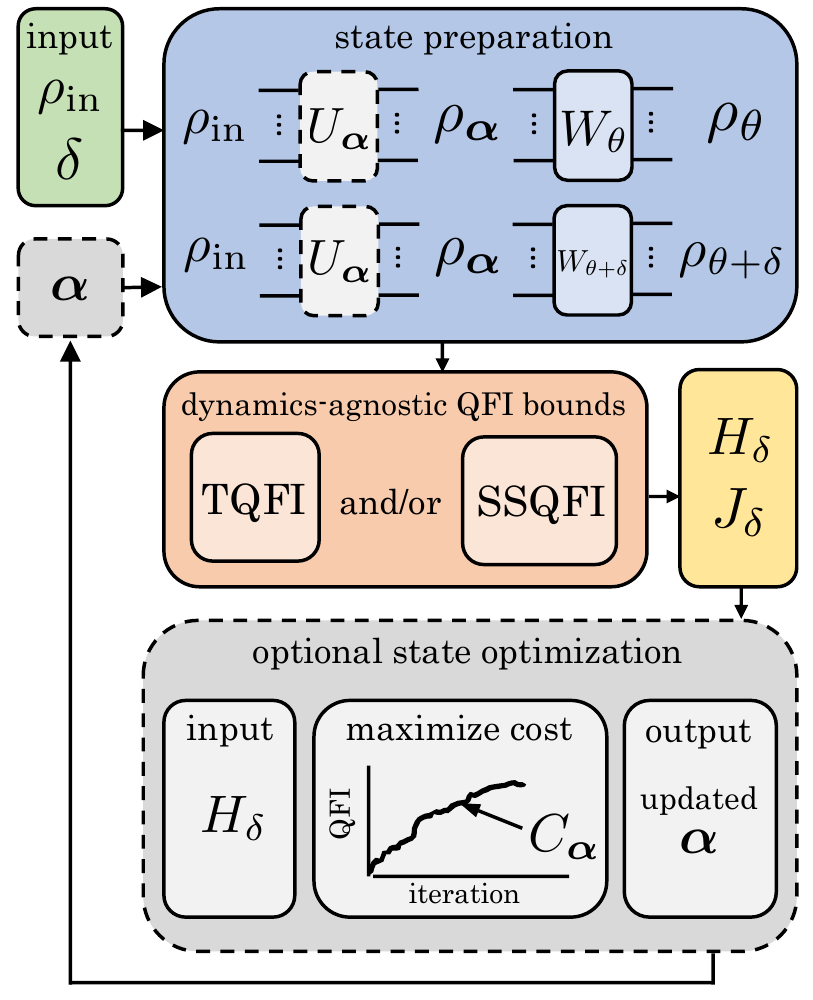}
    \caption{\textbf{Schematic diagram of the VQFIE algorithm for a unitary sensor dynamics application.} VQFIE takes as input $N$ copies of an $n$-qubit state $\rho_{\text{in}}$, a value of $\delta$, and  the parameters $\vec{\alpha}$ of a parameterized unitary $U_{\boldsymbol{\alpha}}$. The computation of the TQFI bound requires $N=2$, while that of the SSQFI requires $N=4$.   After the (optional) application of $U_{\boldsymbol{\alpha}}$, half of the probe states evolve under the action of a unitary $W_{\theta}=e^{- i \theta G}$ as in~\eqref{eq:phase}, while the other half under the action of $W_{\theta+\delta}$, resulting in $N/2$ exact states $\rho_\theta$ and $N/2$ error states $\rho_{\theta+\delta}$. Then, VQFIE computes the dynamic-agnostic bounds $H_\delta$ and $J_\delta$ of Eqs.~\eqref{eq:H_delta}--\eqref{eq:J_delta}. 
    The lower bound $H_\delta$ on the QFI can be employed as a cost function in a quantum-classical hybrid optimization loop to train the parameters $\vec{\alpha}$ and variationally prepare the state that maximizes the QFI.}
    \label{fig:quick-alg}
\end{figure}

In this section, we present a high level description of the VQFIE algorithm, shown in Fig.~\ref{fig:quick-alg}. For completeness, we describe the algorithm including the optional step (dashed boxes in Fig.~\ref{fig:quick-alg}) of variationally preparing the state that maximizes the QFI for estimating the parameter~$\theta$.  

We remark that the VQFIE algorithm is meant to address the situation where the state of interest has a relatively high purity. Physically speaking, this would occur if one attempts to prepare a pure state on an noisy quantum device, which then results in a mixed state with high purity. Low temperature thermal states provide another important example. Specifically, in order for the bounds in VQFIE to remain tight, the input state $\rho_{\text{in}}$ should be approximately low rank, as defined in Ref.~\cite{cerezo2019variational}. We emphasize that such states are of significant physical interest, especially in the context of quantum sensing where one aims to prepare a state with high purity.

We also remark that the task of estimating the QFI for such states (i.e., low-rank mixed states) is likely to be hard for classical computers. This can be seen in a number of ways. First, because the QFI can be expressed in terms of the fidelity between two quantum states, hardness results that apply to fidelity estimation (such as ref. ~\cite{cerezo2019variational}) also apply to QFI. Second, the standard technique for QFI estimation using a classical computer involves posing the problem as a semi-definite program which has computational run time polynomial in the dimension of the states \cite{katariya2021geometric}. Because this dimension scales exponentially with the number of qubits, such techniques are not scalable to problem sizes of interest. As such, a quantum computer could provide advantage in the task of estimating the QFI of an arbitrary mixed quantum state.

\subsection{Algorithm Structure}
\label{sec:VQFIEalgorithm}

\subsubsection{State preparation}

As schematically depicted in Fig.~\ref{fig:quick-alg}, the input to the VQFIE algorithm are $N$ copies of  an $n$-qubit input state $\rho_{\text{in}}$, a value of $\delta$, and a set of parameters $\vec{\alpha}$ which parametrize a  unitary $U_{\vec{\alpha}}$. When computing the TQFI bounds we have $N=2$, whereas the computation of the SSQFI bounds require $N=4$ copies of $\rho_{\text{in}}$. In both cases, we can first apply the parametrized unitary to the input state to obtain a variational probe state
\begin{equation} \label{eq:rho}
    \rho_{\vec{\alpha}}=U_{\vec{\alpha}}\rho_{\text{in}}U\ad_{\vec{\alpha}}\,.
\end{equation} 
The goal of $U_{\vec{\alpha}}$ is to map the input state $\rho_{\text{in}}$ to the state that maximizes the QFI. Note that if one just wants to estimate the QFI, and not variationally maximize it, one need not apply $U_{\boldsymbol{\alpha}}$ to the input state. 

Since no knowledge about the dynamics of the source is assumed, we employ a hardware efficient ansatz~\cite{kandala2017hardware} for the parametrized unitary $U_{\vec{\alpha}}$. This ansatz reduces the circuit depth overhead when implementing VQFIE on a quantum computer by expressing $U_{\vec{\alpha}}$ as a sequence of gates taken from an alphabet of native gates to the specific hardware employed. Hence, without loss of generality we can write
\begin{equation}
    U_{\vec{\alpha}}=\prod_\mu e^{-i \alpha_\mu V_\mu} \Gamma_\mu\,,
\end{equation}
where $\Gamma_\mu$ are unparametrized unitaries, and $V_\mu$ are Hermitian operators. For the numerical implementations on this work we employ a layered hardware efficient ansatz, where gates are arranges in a brick-like structure acting on alternating pairs of neighbouring qubits~\cite{cerezo2020cost}.  Such architecture can be readily implemented in a quantum computer with local qubit connectivity.

After the optional action of $U_{\vec{\alpha}}$, half of the states interact with a source that encodes the information of a parameter $\theta$ as
\begin{equation}\label{eq:phase}
    \rho_{\theta}=W_{\theta}\rho W_{\theta}\ad\quad \text{with}\quad W_{\theta}=e^{-i\theta G}\,,
\end{equation}
while the source encodes the information of $\theta+\delta$ in the remaining half. That is, we obtain $N/2$ exact states $\rho_\theta$ and $N/2$ error states $\rho_{\theta+\delta}$.  Note that, as shown in Fig.~\ref{fig:quick-alg}, our algorithm does not require knowledge of the generator of the interaction between the quantum state and the source, that is, we do not require knowledge about what the Hermitian operator $G$ is. At this stage, it is important to distinguish between $\boldsymbol{\alpha}$ and $\theta$. The former is a set of parameters which can be variationally updated to prepare the near-optimal state for sensing applications, while the later is the parameter one is attempting to estimated (e.g. $\theta$ is the magnetic field amplitude $B$ in the case of magnetometry).

\subsubsection{Computation of bounds}

The next step in VQFIE is the computation of the QFI bounds. As indicated in \eqref{eq:H_delta} and \eqref{eq:J_delta} one needs to compute the TQFI and the SSQFI bounds. Here, we briefly describe how each one of those quantities can be estimated on a quantum computer. 

The SSQFI bounds are obtained by computing the sub- and super-fidelities of Eqs.~\eqref{eqn:subfid}--\eqref{eqn:superfid}. The terms in $E(\rho_{\theta}, \rho_{\theta+\delta})$ and $R(\rho_{\theta}, \rho_{\theta+\delta})$ of the form $\Tr[\rho_{\theta}^2]$ or $\Tr[\rho_{\theta} \rho_{\theta+\delta}]$ can be computed with $2n$ qubits by means of the destructive swap test~\cite{cincio2018learning}. The destructive swap test employs a constant-depth quantum circuit with classical post-processing that scales linear in the number of qubits~\cite{cincio2018learning}. For the $\Tr[\rho_{\theta} \rho_{\theta+\delta}\rho_{\theta} \rho_{\theta+\delta}]$ term, one can employ a generalized swap test (e.g., Refs.~\cite{brun2004measuring,ekert2002direct}) involving a controlled permutation gate, whose circuit depth scales linearly in the number of qubits. For completeness, we show how the generalized swap test can be used to estimate these functionals in Appendix \ref{sec:estimatingFunctionals}. As shown there, these circuits are efficient in the problem size. Finally, we remark that one can also compute the SSQFI bounds via the circuits introduced in~\cite{MiszczakPHUZ09}.

Computing the TQFI bounds is a more involved procedure and requires a variational subroutine. Specifically, we will need to obtain the $m$ largest eigenvalues and associated eigenvectors of $\rho_{\theta}$. These are obtained using the Variational Quantum State Eigensolver algorithm~\cite{cerezo2020variational}, which variationally diagonalizes the state $\rho_{\theta}$ over the subspace of its $m$ principal components. Specifically, one trains a parametrized gate sequence to achieve this subspace diagonalization task. 
The subroutine then returns estimates of the $m$ largest eigenvalues and their associated eigenvectors, denoted $\{\tilde{\lambda}_i\}_{i=1}^m$ and $\{ \ket{\tilde{\lambda}_i}\}_{i=1}^m$, respectively. We refer the reader to Appendix  \ref{sec:VQFIEalgorithmdetail}  for additional details on these subroutines.

After this variational subroutine, one then runs several non-variational quantum circuits to compute the overlap between $\rho_{\theta+\delta}$ and the estimates of the principal components of $\rho_{\theta}$, i.e., the states in the set $\{ \ket{\tilde{\lambda}_i}\}_{i=1}^m$. These overlaps are then combined with classical post-processing as described in Ref.~\cite{cerezo2019variational}, in order to compute the upper and lower bounds on the fidelity appearing in \eqref{eq:TFB1}.  We remark that Ref.~\cite{cerezo2019variational} showed that this procedure scales efficiently with problem size. Here we remark that the efficiency in estimating the TQFI bounds relies on the efficiency of the variational diagonalization subroutine~\cite{cerezo2020variational}.

\subsubsection{Classical Parameter Update}
When preparing the optimal probe state $\rho_{\vec{\alpha}}$ the final step of each VQFIE iteration is a classical parameter update. Here our algorithm learns the parameters $\vec{\alpha}$ that approximately maximize the cost function
\begin{align}\label{eq:cost}
 C_{\vec{\alpha}}  = H_{\delta}(\theta;\rho_{\theta})\,,
\end{align}
where we note that the dependence on $\vec{\alpha}$ that arises from the preparation unitary is left implicit to simplify the notation. Here, a hybrid quantum-classical optimizer employs the value of the cost (or its gradient) to update the preparation parameters, $\vec{\alpha}$. The whole algorithm then repeats until stopping criteria are met. The probe state from the final iteration, which approximately maximizes Eq.~\eqref{eq:cost}, is then used to calculate the upper bounds. The estimation of the QFI is then between $H_{\delta}$ and $J_{\delta}$.

\subsection{Gradient scaling}
\label{sec:gradientscaling}

Significant progress has recently been made on studying the scaling of gradients in VQAs~\cite{mcclean2018barren, cerezo2020cost,sharma2020trainability,volkoff2020large,wang2020noise,cerezo2020impact}. This includes identifying some conditions under which the gradient vanishes exponentially in $n$, known as a barren plateau landscape.

We now proceed to argue that VQFIE does not exhibit a barren plateau landscape when a shallow-depth ansatz is employed. In particular, here we will use the results in~\cite{cerezo2020cost} that connect the existence of barren plateaus in VQAs to the locality of the cost function. Specifically, it was shown that global cost functions, i.e. cost functions where one computes the expectation value of operators acting non-trivially on all qubits, lead to barren plateaus. On the other hand, local cost functions, i.e., cost functions where one computes the expectation value of operators acting non-trivially on a small subset of qubits, do not exhibit barren plateaus for shallow ansatzes~\cite{cerezo2020cost}. Hence we aim to argue that the cost functions employed in VQFIE are local, rather than global, in nature.

First we consider the variational subroutine used to compute the TQFI bounds. This is the Varitaional Quantum State Eigensolver, proposed in Ref.~\cite{cerezo2020variational}, and therein a local cost function was proposed for the diagonalization task of interest. We refer the reader to Ref.~\cite{cerezo2020variational} for details on this local cost function. But it suffices to say that the cost landscape would not have a barren plateau so long as the ansatz depth is sufficiently shallow, i.e., $\OC(\log (n))$ depth~\cite{cerezo2020cost}.

\begin{figure*}[htp!]
\centering
\includegraphics[width=0.95\textwidth]{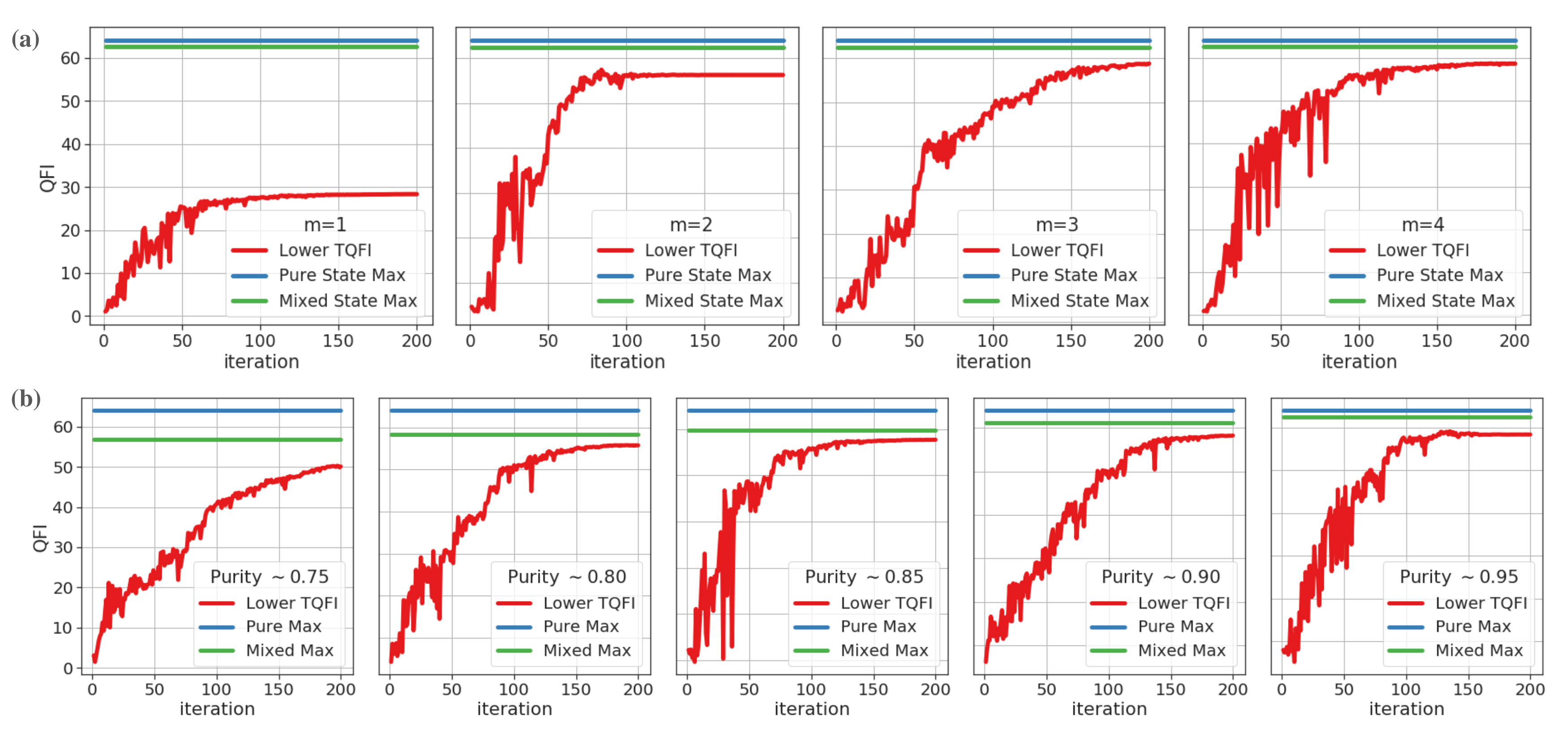}
\caption{\textbf{Cost function value versus iteration for an $n=4$ implementation of VQFIE.} (a) Panels correspond to a different value of $m$ for the TQFI lower bound. The input state was  randomly generated with a purity  of $0.95$ and the  error state was created with $\delta=0.1$. In all cases the TQFI is tighter than the SSQFI. The plots show that  the performance of VQFIE improves as $m$ increases since VQFIE reaches higher cost values.  (b) Panels correspond to randomly generated states with different purities. In all cases the error state was created with $\delta=0.1$. We see that VQFIE reaches higher cost values for higher purity states. Horizontal lines depict the maximal QFI for pure states (Eq.~\eqref{eq:maxpurestate}) and for mixed states (Eq.~\eqref{eq:maxmixedstate}). }
\label{fig:cost-v-iteration}
\end{figure*}

Second we consider the variational optimization of the $\vec{\alpha}$ parameters in the state preparation unitary. This involves the cost function in \eqref{eq:cost}, which takes the maximum between the lower bounds provided by the TQFI and the SSQFI. In our implementation of VQFIE, we found the TQFI to typically provide a tighter bound~\cite{cerezo2019variational}, so we focus our discussion here on the TQFI. From Ref.~\cite{Sone2020QFI}, we can write the TQFI lower bound in the limit $\delta\rightarrow 0$ as

\begin{align}
\begin{split}
\label{eq:TQFIxpression}
 \IC\left(F_{*}; \rho_{\theta}^{(m)}\right)
 =&4 \Tr\left[\rho^{(m)}_{\theta} G^2\right]-\sum_{i,j=1}^m\frac{8\lambda_i\lambda_j}{\lambda_i+\lambda_j}|G_{ij}|^2\\
 &-4 \Tr\left[\Pi^{\overline{m}}_{\rho_{\theta}} G  \rho^{(m)}_{\theta} G\right],
\end{split}
\end{align}
with $G_{ij}=\bramatket{\lambda_i}{G}{\lambda_j}$, and where  $\Pi^{m}_{\rho_{\theta}}+\Pi^{\overline{m}}_{\rho_{\theta}}=\id$ .

In this form, we see that the first term is a local cost function~\cite{cerezo2020cost}, so long as  $G$ is local, and hence its gradient should remain large for a shallow depth ansatz. For instance, for a magnetometry application (see below) one has  $G=\sum_{i=1}^n Z_i$, and hence $G^2=n\id +\sum_{i\neq j} Z_iZ_j$ contains terms that act non-trivially on at most two qubits. In contrast, the last term is a global cost function. Recalling that  $\Pi^{\overline{m}}_{\rho_{\theta}}=\sum_{k=m+1}^d\dya{\lambda_k}$, the final term in~\eqref{eq:TQFIxpression} is of the form $\sum_{k=m+1}^d\Tr\left[G\dya{\lambda_k} G  \rho^{(m)}_{\theta}\right]$, and one computes the expectation value of the global operators $G\dya{\lambda_k} G$. (Note that this term is global regardless of the locality of $G$.) Hence, the contribution to the cost function gradient associated with this final term will be exponentially suppressed. Finally, we note that the middle term is likely to have a smaller gradient than the first term even though it is not a fully global term. Hence, we expect that the first term in this expression will have the largest gradient magnitude, and as a result the overall gradient magnitude will not vanish exponentially for an $\OC(\log (n))$ depth ansatz.

To further support the claims that our algorithm has a local cost function, and hence does not exhibit a barren plateau for shallow ansatzes, we refer the reader to the numerical simulation section. Therein we numerically analyze the trainability of our cost function.

\section{Numerical simulation}
\label{sec:numericalsimulation}

In this section, we present our numerical results obtained from simulating the VQFIE algorithm. Specifically, we train the parameters in $U_{\vec{\alpha}}$ in order to prepare the probe state that maximizes the QFI for a magnetometry application. Hence, we consider a system of $n$ spin-1/2 particles ($n$ qubits) interacting with a uniform magnetic field. The Hamiltonian is modeled as
\begin{equation}\label{eq:hammag}
    G=\sum_{i=1}^n Z_i\,,
\end{equation}
with $Z_i$ the Pauli $z$ operator on qubit $i$. The parameter $\theta$ appearing in Eq.~\eqref{eq:phase} is the phase acquired by spins, after precessing for some time under the action of the magnetic field.

Here we recall that if the probe  state is pure $\rho_{\vec{\alpha}}=\dya{\psi_{\vec{\alpha}}}$, then it is well known that the optimal probe state corresponds to the GHZ state $\ket{\GHZ}=(\ket{0}^{\otimes n}+e^{i\varphi}\ket{1}^{\otimes n})/\sqrt{2}$ with $\varphi\in\Rbb$, and the QFI reaches the Heisenberg limit
\begin{equation}\label{eq:maxpurestate}
 \max_{\ket{\psi_{\vec{\alpha}}}}   I(\theta,\dya{\psi_\theta})=4n^2\,.
\end{equation}
Here $\ket{\psi_\theta}=W_\theta \ket{\psi_{\vec{\alpha}}} $. Moreover, if the probe state $\rho_{\vec{\alpha}}$ is mixed, then the optimal state can be obtained from Ref.~\cite{fiderer2019maximal}, and its associated QFI is
\begin{equation}\label{eq:maxmixedstate}
    \max_{\rho_{\vec{\alpha}}} I(\theta,\rho_\theta)= \frac{1}{2}\sum_{k=1}^{d}\lambda_{k,d-k+1} (g_k-g_{d-k+1})^2\,,
\end{equation}
where $g_k$ are the eigenvalues of $G$ ordered in decreasing order and where $d=2^n$. Here, $\lambda_{k,l}=0$ if $\lambda_k=\lambda_l=0$ and $\lambda_{k,l}=(\lambda_k-\lambda_l)^2/(\lambda_k+\lambda_l)$ otherwise~\cite{fiderer2019maximal}, and we recall that $\lambda_k$ is the $k$-th eigenvalue of the probe state $\rho_{\vec{\alpha}}$. Note, that in order to compute~\eqref{eq:maxmixedstate} one must have perfect knowledge of all the eigenvalues of $\rho_{\vec{\alpha}}$ and of the generator $G$. Hence, such quantity is not efficiently computable in practice.  However, here we employ Eqs.~\eqref{eq:maxpurestate}--\eqref{eq:maxmixedstate} to benchmark the maximum QFI obtained by training the parameters in $U_{\vec{\alpha}}$. 

\subsubsection{Performance of VQFIE in experimentally relevant regimes}
For our heuristics we simulated the VQFIE algorithm without sampling noise. Moreover, the cost function optimization was performed by employing the Constrained
Optimization By Linear Approximation (COBYLA) algorithm~\cite{PowellCOBYLA}. For each case analyzed we ran 30 instances of VQFIE, each with $200$ cost optimization iterations, and we present the results of the run that achieved the largest final cost function value.  We remark that for $U_{\vec{\alpha}}$ we employed a layered hardware efficient ansatz with three layers composed of single qubits rotations and CNOT gates.  

In Fig.~\ref{fig:cost-v-iteration}(a) we show results for an $n=4$ qubit implementation of the VQFIE algorithm for a randomly generated mixed state with a purity of $0.95$ and for different values of $m$ in the truncated state $\rho_\theta^{(m)}$.  In all cases, the TQFI lower bound was tighter than the SSQFI lower bound so that $H_{\delta}(\theta;\rho_{\theta}) = \IC_\delta(F_{*},\rho_\theta^{(m)})$. Here we can verify that the TQFI lower bound becomes tighter with increasing $m$. Note that the improvement in the $m=4$ case is not as significant due to the fact that smaller eigenvalues give rise to smaller improvements. This is due to the fact that most of the information in the state is encoded in the subspace spanned by the eigenvectors associated with the largest eigenvalues~\cite{cerezo2019variational}.

Let us now analyze the performance of VQFIE for different purities. In Fig.~\ref{fig:cost-v-iteration}(b)  we present results for randomly generated $n=4$ input states with purities of $0.75,0.8,\ldots,0.95$. Let us first remark that the maximum QFI achievable increases with the purity, as shown by the vertical lines obtained from Eq.~\eqref{eq:maxmixedstate}. In all cases we chose $m=4$, and for all purities we found that the TQFI lower bound is tighter than the SSQFI lower bound.  Moreover, we can also see that, as expected, VQFIE has a better performance for high purity states, as the final cost value is larger for higher purities.  This again can be explained from the fact that in low purity states, more information is encoded in a larger number of eigenvalues and the subspace generated by their associated eigenvectors. Hence, truncating the state to $m=4$ leads to information loss and to looser truncated bounds. Similarly, the SSQFI bounds are also loose for low purities due to the looseness of the sub- and super-fidelities in this case~\cite{cerezo2019variational}.

We note that, for quantum metrology applications, one  wishes  to prepare probe states that are typically low rank (high purity), as states that are close to being maximally mixed are not good candidates for quantum sensors~\cite{fiderer2019maximal}. Hence, it is fortunate that our algorithm has a better performance in this high-purity regime.

\begin{figure*}[t]
  \includegraphics[width=1.00\textwidth]{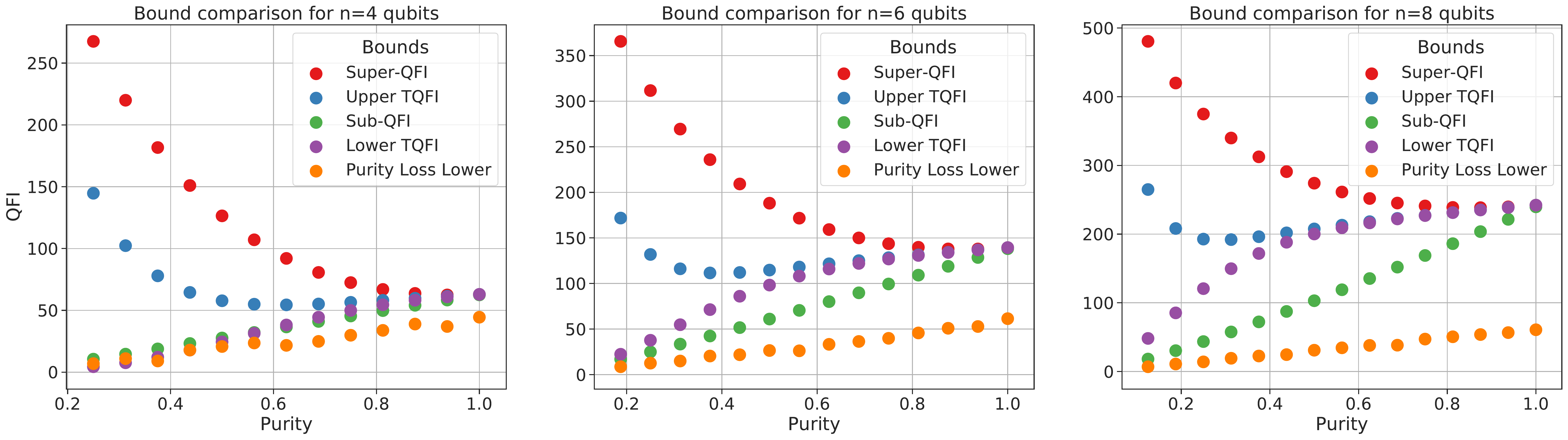}
  \caption{\textbf{VQFIE and purity loss bounds.} Here we present the TQFI, SSQFI and the purity loss bounds on the QFI for different system sizes ($n=4,6,8$) and for purities in $(1/n,1)$. For each value of the purity the bounds were evaluated at the optimal probe state of~\cite{fiderer2019maximal} which maximizes the mixed state QFI in Eq.~\eqref{eq:maxmixedstate}. Here we can see that in all cases considered the VQFIE lower bounds of~\eqref{eq:H_delta} is tighter than the purity loss bound in~\eqref{eq:purityloss2}. Moreover, the plots also show that the VQFIE bounds in Eq.~\eqref{eq:VQFIEbounds} are tighter for high purities, i.e., in the range of purities for which VQFIE is aimed. }
  \label{fig:BoundComparison}
\end{figure*}

\subsubsection{Scaling of the cost function landscape for VQFIE}

In this section we numerically analyze the presence or absence of barren plateaus in the VQFIE cost function landscape. Specifically, we consider here the case when one trains the parameters $\vec{\alpha}$ via the TQFI lower bound. In Section~\ref{sec:gradientscaling}, we argued that in this case the cost function is local, and hence that no barren plateau arises for shallow hardware efficient ansatzes (i.e., with $\OC(\log(n))$ depth). Here we recall that the standard way to numerically analyze the existence of barren plateaus is by computing the variance of the cost function partial derivative, leading to the following definition.
\begin{definition}
A cost function $C_{\vec{\alpha}}$ exhibits a barren plateau if $E_{\vec{\alpha}}[\partial_\mu C_{\vec{\alpha}}/n^2]=0$ and if 
\begin{equation}\label{eq:BP}
    \Var_{\vec{\alpha}}[\partial_\mu C_{\vec{\alpha}}/n^2]\in\OC(1/2^n)\,,
\end{equation}
where $\partial_\mu C_{\vec{\alpha}}\equiv \partial C_{\vec{\alpha}}/\partial{\alpha_\mu}$ for some $\alpha_\mu\in\vec{\alpha}$.
\end{definition}
Here the expectation values are taken over the parameters $\vec{\alpha}$, and we divided the cost function by the normalization factor $n^2$. Note that~\eqref{eq:BP} implies that the cost function gradients are (on average) exponentially suppressed across the landscape, and hence that the landscape is essentially flat for large problem sizes.

Recently, it was shown in~\cite{arrasmith2021equivalence} that the presence of barren plateaus can also be diagnosed via the variance of difference in cost function values, i.e., by analyzing the scaling of $ \Var_{\vec{\alpha},\vec{\alpha}'}[\Delta C_{\vec{\alpha},\vec{\alpha}'}/n^2]$, where $\Delta C= C_{\vec{\alpha}}-C_{\vec{\alpha}'}$. The main advantage here is that $ \Var_{\vec{\alpha},\vec{\alpha}'}[\Delta C_{\vec{\alpha},\vec{\alpha}'}/n^2]$  is computationally cheaper to compute, as it requires less quantum circuit evaluations~\cite{arrasmith2021equivalence}.  Hence we have the following alternative definition.
\begin{definition}
A cost function $C_{\vec{\alpha}}$ exhibits a barren plateau if $E_{\vec{\alpha}}[\partial_\mu C_{\vec{\alpha}}/n^2]=0$ and if 
\begin{equation}\label{eq:BP_2}
   \Var_{\vec{\alpha},\vec{\alpha}'}[\Delta C_{\vec{\alpha},\vec{\alpha}'}/n^2]\in\OC(1/2^n)\,.
\end{equation}
\end{definition}

\begin{figure}[htp!]
\includegraphics[width=0.5\textwidth]{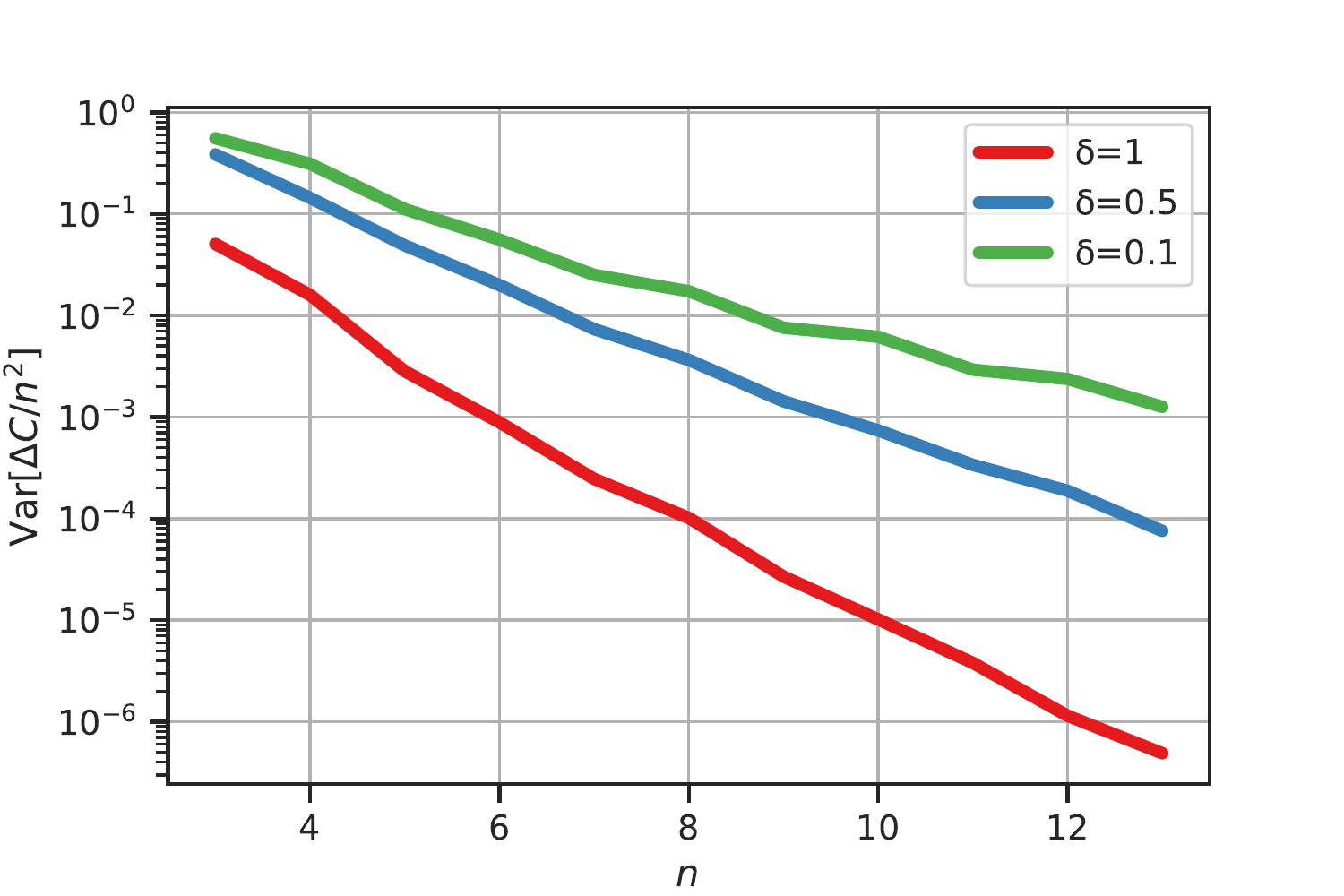}
\caption{\textbf{Normalized variance of the VQFIE cost function difference versus number of qubits.} As shown in Ref.~\cite{arrasmith2021equivalence}, the y-axis can viewed as a proxy for the trainability of the cost function. If the variance of the cost difference vanishes exponentially with the number of qubits, the cost exhibits a barren plateau. However, for $\delta=0.1$ we see that this variance does not decrease exponentially with problem size. Namely, since the $y$-axis is presented in a log scale, and since the green curve deviates significantly from a straight line, then the scaling is non-exponential.}
\label{fig:delta-v-n}
\end{figure}

In Fig.~\ref{fig:delta-v-n}, we show numerical results for $\Var_{\vec{\alpha},\vec{\alpha}'}[\Delta C_{\vec{\alpha},\vec{\alpha}'}/n^2]$ versus the number of qubits ($n=2,3,\ldots,13$) for the VQFIE cost function defined in Eq.~\eqref{eq:cost}. We employ a hardware efficient ansatz with $\log(n)$ layers for the magnetometry application of Eq.~\eqref{eq:hammag}, with different values of $\delta=0.1,0.5,1$. For $n=2,\ldots,11$ we computed by variance by averaging over $1000$ random parameter initializations, while for $n=12,13$ we averaged over $100$ initializations.  Here we can we see that for $\delta=1$ the variance of cost function differences vanishes exponentially with the system size (as noted by a straight line in the log-linear plot). In this case, the cost function is clearly global as one compares two states in an exponentially large Hilbert space whose fidelity is (on average) exponentially small. Such case is similar to that analyzed in~\cite{cerezo2020cost}, where a cost function defined as  the overlap between two quantum states that are not necessarily close exhibits a barren plateau.

On the other hand, as $\delta$ decreases the scaling of $\Var_{\vec{\alpha},\vec{\alpha}'}[\Delta C_{\vec{\alpha},\vec{\alpha}'}/n^2]$  drastically improves since the cost becomes local and Eq.~\eqref{eq:TQFIxpression} holds. Specifically, for $\delta=0.1$ the variance $\Var_{\vec{\alpha},\vec{\alpha}'}[\Delta C_{\vec{\alpha},\vec{\alpha}'}/n^2]$ clearly does not vanish exponentially with $n$, which means that the cost function does not exhibit a barren plateau. This non-exponential scaling is characteristic of  local cost functions~\cite{cerezo2020cost}. Moreover, one can see that $\Var_{\vec{\alpha},\vec{\alpha}'}[\Delta C_{\vec{\alpha},\vec{\alpha}'}/n^2]$ is increased by orders of magnitude by reducing the value of $\delta$. Taken together, these results show that in the small delta limit, the cost function trainability is improved as the cost function becomes local.

\section{Comparison to Literature} \label{sec:lowerboundcomparison}

In this section we heuristically compare the TQFI and the SSQFI lower  bounds with the so-called purity loss bound, a dynamics-agnostic lower bound for the QFI ~\cite{Modi16,Yang20}. As shown below, the bounds presented in this manuscript are tighter than the purity loss bound for all cases considered.

As shown in~\cite{Modi16,Yang20}, the following lower bound on the QFI holds
\begin{align}\label{eq:puritylower}
\LC(\theta;\rho_{\theta})\leq\IC\left(F_{*};\rho_{\theta}^{(m^{*})}\right)\,,
\end{align}
where  $\LC(\theta;\rho_{\theta})= 4\left(\Tr\left[\rho^2 G^2\right]-\Tr\left[\rho G \rho  G\right]\right)$. However, this quantity can also be expressed as~\cite{Modi16}
\begin{align}\label{eq:purityloss2}
    \LC(\theta;\rho_{\theta})\approx 2\frac{\Delta\nu}{\left(\Delta x\right)^2}\,,
\end{align}
 where $\Delta\nu$ is the purity loss given by
\begin{align}
\Delta\nu = \Tr\left[\rho^2\right]-\Tr\left[\rho_{\text{ave}}^2\right]\,.
\end{align}
Here, $\rho_{\text{ave}}$ is the the ensemble-averaged state obtained by considering that the parameter $\theta$ is not stable during an experiment but rather is subject to statistical fluctuations. Hence, each time the probe states interacts with the source the unitary $W_{\theta(x)}$ is applied with some probability $p_x$. Finally, $\theta(x)$ is a random variable normally distributed about $\theta$ with some variance $(\Delta x)^2\ll 1$, and we define  $\rho_{\text{ave}}=\sum_x p_x \rho_{\theta(x)}$~\cite{Modi16}. Hence,  $\LC(\theta;\rho_{\theta})$ quantifies how fragile the probe state is to stochastic fluctuations in the parameter $\theta$.

In Fig.~\ref{fig:BoundComparison} we compare the TQFI and the SSQFI bounds with the purity loss bound of~\eqref{eq:purityloss2} for different system sizes ($n=4,6,8$) and for states with purities in the range $(1/n,1)$. In all cases, the probe state was the mixed state that leads to the maximized QFI of~\eqref{eq:maxmixedstate} from~\cite{fiderer2019maximal}. Just as with the shift parameter $\delta$ in Eq. \eqref{eq:fidelityQFI}, one would ideally minimize $(\Delta x)^2$; however, achievable ranges depend on experimental capabilities. So to fairly compare to Ref.~\cite{Yang20}, we set $\delta=(\Delta x)^2=0.1$, as this was the value used in their experiment. Here we remark that the TQFI bounds were computed with $m=4$ while the purity of the average state $\rho_{\text{ave}}$ was obtained by stratified sampling from a discretized Gaussian distribution with $K$ samples (or strata)~\cite{Yang20}. We refer the reader to Appendix~\ref{sec:comparison} for a detailed discussion of how the number of strata was obtained for a fair comparison between bounds. 

As shown in Fig.~\ref{fig:BoundComparison}, the VQFIE lower bound of~\eqref{eq:H_delta} is tighter than the purity loss bound for all values of $n$ and purities considered. In fact, both the TQFI and the SSQFI lower bounds are individually tighter than the purity loss bound, with the only exception being $n=4$.  Here $\LC(\theta;\rho_{\theta})$ is larger than the TQFI lower bound for purities smaller than 1/2. As previously mentioned, this is expected due to fact that the TQFI bounds are loose for low purity states. Moreover, we can also see that for $n=6$ and $n=8$ the VQFIE bounds are noticeably better than the purity loss bounds. This can be due to the fact that for~\eqref{eq:purityloss2} to hold one requires $(\Delta x)^2\ll 1$, which is not always the case~\cite{Yang20}. 

Finally, let us remark that Fig.~\ref{fig:BoundComparison} also shows that the VQFIE upper and lower bounds can be very tight for high purities. This means that  $J_\delta-H_\delta$ will give a small interval where the QFI actually lies, and hence VQFIE outputs a precise estimate of the QFI in this purity range.

\section{Conclusion}
\label{sec:discussion}

In this work, we presented an algorithm designed for NISQ devices to estimate the quantum Fisher information (QFI), called Variational Quantum Fisher Information Estimation (VQFIE). For this purpose, we introduced new upper and lower bounds on the QFI that are based on bounding the fidelity. These bounds are then efficiently computed on a quantum computer, as part of our proposed algorithm.

Specifically, we presented two types of bounds on the QFI that are conceptually distinct, in that while both attempt to tackle the problem of computing the quantum fidelity (a non-linear function of quantum states), the approach used is different in each case. One of our bounds is based on truncating the spectrum of the exact state and reducing the complexity of computing the non-linear function on (sub-normalized) quantum states. These are are called the Truncated Quantum Fisher Information (TQFI) bounds. Our other bounds are based on bounds for the quantum fidelity known as the  Sub- and Super-fidelity, which replace the non-linear function in the fidelity by linear or quadratic functions.  In this case, the bounds are called  the  Sub- and Super-Quantum Fisher Information (SSQFI) bounds. 

We especially focused on the TQFI bounds. Our previous work established the TQFI lower bound as being operationally meaningful~\cite{Sone2020QFI}.  The present paper focuses on computing the TQFI bounds using a variational quantum algorithm, and we show how the maximization of the lower bound over state preparations can be used to enhance quantum sensing capabilities. 

Although we found the TQFI to be tighter for the magnetometry Hamiltonian considered, in general this may not be the case. As such, we include the possibility in our algorithm to include both TQFI and SSQFI bounds, keeping only the tightest in the end. Although the TQFI bounds were tighter in our specific implementation, there were two distinct benefits to investigating the SSQFI bounds. First, in a companion paper~\cite{cerezo2021sub}, we proved that the Sub-QFI is a faithful lower bound, meaning that the state that maximizes the sub-QFI is the same state that maximizes the QFI. Second the SSQFI bounds do not require the variational quantum subroutine to truncate the state to the principal components. 

In addition to introducing the general algorithm, we provided qualitative and quantitative arguments as to why our algorithm avoids barren plateaus for shallow depth ansatzes and for small $\delta$ values, suggesting a favorable scaling for the gradient magnitude, and hence a favorable scaling for training.

While trainability is a crucial consideration when proposing lower bounds for use in VQA cost functions, so too is the tightness of the proposed bounds. As illustrated in our numerical results, we expect our bounds to be tighter as the purity of the state increases. This is also true for other bounds on the QFI in the literature. For the magnetometry example that we considered, we found that our bounds were tighter than recently proposed literature bounds, over a range of different purity values.

In addition to tightness and trainability, another key aspect of VQFIE is the fact that our bounds are agnostic to the underlying dynamics. Computing our bounds does not require knowledge of the generator of the dynamics. This is useful for quantum sensing tasks for systems that are either complicated or not fully characterized. Hence, this makes the VQFIE algorithm broadly applicable.

VQFIE is a promising algorithm for implementation in the NISQ era. Because the  quantum Cram{\'{e}}r-Rao bound (QCRB)  is ubiquitously used as a figure of merit in experiments, efficiently and accurately estimating the QFI (upon which the QCRB directly depends) is a crucially important task in quantum sensing. As such, we expect that our algorithm will find broad applicability in evaluating the performance of quantum sensors in the fields of chemistry, biology, material science, and cosmology.

\bigskip
     
\section*{Acknowledgements}

We are grateful to Tyler Volkoff, Jing Liu, and Haidong Yuan for helpful discussions. This work was supported by the Quantum Science Center (QSC), a National Quantum Information Science Research Center of the U.S. Department of Energy (DOE). JLB was initially supported by the U.S. DOE through a quantum computing program sponsored by the LANL Information Science \& Technology Institute. JLB was also supported by the National Science Foundation Graduate Research Fellowship under Grant No. 1650115.  MC and AS also acknowledge initial support from the Center for Nonlinear Studies at Los Alamos National Laboratory (LANL). AS is now supported by the internal R\&D from Aliro Technologies, Inc.  PJC also acknowledges initial support from the LANL ASC Beyond Moore's Law project.

\bibliography{ref.bib}

\newpage

\onecolumngrid

\pagebreak

\setcounter{section}{0}
\setcounter{proposition}{0}
\setcounter{theorem}{0}
\setcounter{corollary}{0}

\section*{Appendix}

\section{Details on the VQFIE algorithm}

\label{sec:VQFIEalgorithmdetail}

This appendix aims to make our paper more self-contained by providing additional details on the different VQFIE sub-routines employed to compute the TQFI bounds. However, we also refer the reader to the original papers on state diagonalization~\cite{cerezo2020variational} and fidelity estimation~\cite{cerezo2019variational}.

\subsubsection*{State Diagonalization}
Let us here describe the Variational Quantum State Eigensolver (VQSE) algorithm of ~\cite{cerezo2020variational}. As described in the main text, VQSE is employed to obtain approximates of the $m$-largest eigenvalues of a state, and to prepare their associated approximated eigenvectors. Here we recall that we use this algorithm as a subroutine to compute the TQFI as schematically shown in  Fig. \ref{fig:quick-alg}.

The VQSE algorithm takes in as input an integer $m$, a quantum state $\rho$, and a set of parameters $\vec{\beta}$ used to parameterize a diagonalizing gate sequence which we denote as $V_{\vec{\beta}}$.  Then, the output are estimates of the $m$ largest eigenvalues and their associated eigenvectors. Here, the algorithm aims to minimize a  cost function of the form
\begin{align}\label{eq:VQSE-cost}
    C_{\vec{\beta}} = \Tr{\left[H V_{\vec{\beta}} \rho V_{\vec{\beta}}\ad\right]}\,,
\end{align}
for some Hamilitonian $H$ diagonal in the computational basis such that its  $m$ lowest eigenenergies are non-degenerate. This cost function exploits the close connection between majorization and diagonalization, as the cost is minimized if $V_{\vec{\beta}}$ maps the $k$-th largest eigenvector of $\rho$ to the $k$-th smallest energy eigenstate of $H$. 

The parameters in  $\vec{\beta}$ are trained in a hybrid quantum-classical optimization loop whose trainability is guaranteed from the fact that one can always choose $H$ to be a local Hamiltonian~\cite{cerezo2020cost}. Once the optimal parameters $\vec{\beta}$ have been obtained it is then straightforward to extract approximates of the largest eigenvalues $\{\widetilde{\lambda}_i\}_{i=1}^m$ and their associated eigenvectors $\{\ket{\widetilde{\lambda}_i}\}_{i=1}^m$. To estimate the $m$ largest eigenvalues, one simply acts the optimal gate sequence, $V_{\vec{\beta}}$, on $\rho_{\theta}$ and then measures in the computational basis. Mathematically, we have 
\begin{align} \label{eq:approx-vals}
    \tilde{\lambda}_i = \bra{z_i} V_{\vec{\beta}_{\text{opt}}}\rho_{\theta} V_{\vec{\beta}_{\text{opt}}}\ad \ket{z_i}\,,
\end{align}
and in practice, one simply measures the approximately diagonalized state a finite number of times. We denote this number as $N_{\text{runs}}$. Then, for a bitstring $z_i$ with frequency of occurrence $f_i$, the eigenvalues are estimated as
\begin{align}
    \tilde{\lambda}_i \approx \frac{f_i}{N_{\text{runs}}}.
\end{align}
Once the $m$ largest eigenvalues are approximated, the associated eigenvectors can be obtained via
\begin{align}\label{eq:approx-vecs}
    \ket{\tilde{\lambda}_i} &= V_{\vec{\beta}_{\text{opt}}}\ad \ket{z_i},
\end{align}
where $\ket{z_i}=X^{z_i} \otimes \dots \otimes X^{z_n} \ket{0}^{\otimes n}$ and were  $z_i \in \{0,1\}$.

\subsubsection*{Computing TQFI Bounds}

In this section we describe the Variational Quantum Fidelity Estimation algorithm (VQFE) of~\cite{cerezo2019variational}. This algorithm is employed as a subroutine in VQFIE to estimate the TQFI bounds. 

The input to VQFE is an $n$-qubit state $\rho_{\theta+\delta}$ and the estimates of the $m$ largest eigenvalues  $(\{\tilde{\lambda}_i\}_{i=1}^m)$  and associated eigenvectors $(\{ \ket{\tilde{\lambda}_i}\}_{i=1}^m)$ of  $\rho_{\theta}$, both of which are obtained  from VQSE. The goal of VQFE is to compute the generalized fidelity, which we recall for convenience:
\begin{equation}
    F_*(\rho_{\theta}^{(m)},\rho_{\theta+\delta}^{(m)}) = \norm{\sqrt{\rho_{\theta}^{(m)}} \sqrt{\rho_{\theta+\delta}^{(m)}}}_1+ \sqrt{(1-\Tr{[\rho_{\theta}^{(m)}]})(1-\Tr{[\rho_{\theta+\delta}^{(m)}]})}\,.
\end{equation}
Here, we  define the so called $T$-matrix, whose elements are given by 
\begin{align}\label{eq:T-matrix}
    T_{ij} = \sqrt{\tilde{\lambda}_i \tilde{\lambda}_j} \left(\rho_{\theta+\delta}^{(m)}\right)_{ij}
\end{align}
where 
\begin{align}
    (\rho_{\theta+\delta}^{(m)})_{ij} = \bra{\tilde{\lambda}_i} \rho_{\theta+\delta}^{(m)} \ket{\tilde{\lambda}_j}\,.
\end{align}
Note that the first term in the generalized fidelity can be expressed  as
\begin{align}\label{eq:fidelity-via-T}
    \norm{\sqrt{\rho_{\theta}^{(m)}} \sqrt{\rho_{\theta+\delta}^{(m)}}}_1 = \Tr{\sqrt{\sum_{i,j}T_{ij} \ket{\tilde{\lambda}_i} \bra{\tilde{\lambda}_j}}}
    \,.
\end{align}
As described in~\cite{cerezo2019variational}, since we have access to the circuit $V_{\vec{\beta}}\ad$ that prepares the estimated eigenvectors $\ket{\widetilde{\lambda}_i}$, then the matrix elements   $(\rho_{\theta+\delta}^{(m)})_{ij}$  can be efficiently estimated in a quantum computer via a non-variational algorithm. Hence, with post-processing one can always classically create and diagonalize the $m\times m$ $T$-matrix to obtain~\eqref{eq:fidelity-via-T}. This step is efficient as we assume that $m\in\OC(\poly(n))$.  Similarly, it is straightforward to see that the second term in the generalized fidelity is also completely determined by $\{\tilde{\lambda}_i\}_{i=1}^m$ and by  $(\rho_{\theta+\delta}^{(m)})_{ij}$ (both of which are  known $\forall i,j=1,\ldots,m$) as 
\begin{align}
\begin{split}\label{eq:gen-fidelity-simplified}
    F_{*}(\rho_{\theta}^{(m)},\rho_{\theta+\delta}^{(m)}) = \Tr{\sqrt{\sum_{i,j}T_{ij} \ket{\tilde{\lambda}_i} \bra{\tilde{\lambda}_j}}} + \sqrt{\left(1-\sum_i \tilde{\lambda}_i \right)\left(1-\sum_i (\rho_{\theta+\delta}^{(m)})_{ii}\right)}\,.
\end{split}
\end{align}
Hence, both TQFI bounds can be obtained from the (known) terms in~\eqref{eq:gen-fidelity-simplified}. 

\subsubsection*{Computing SSQFI Bounds}\label{sec:estimatingFunctionals}

\begin{figure}[h]
    \centering
    \includegraphics[width=0.5\textwidth]{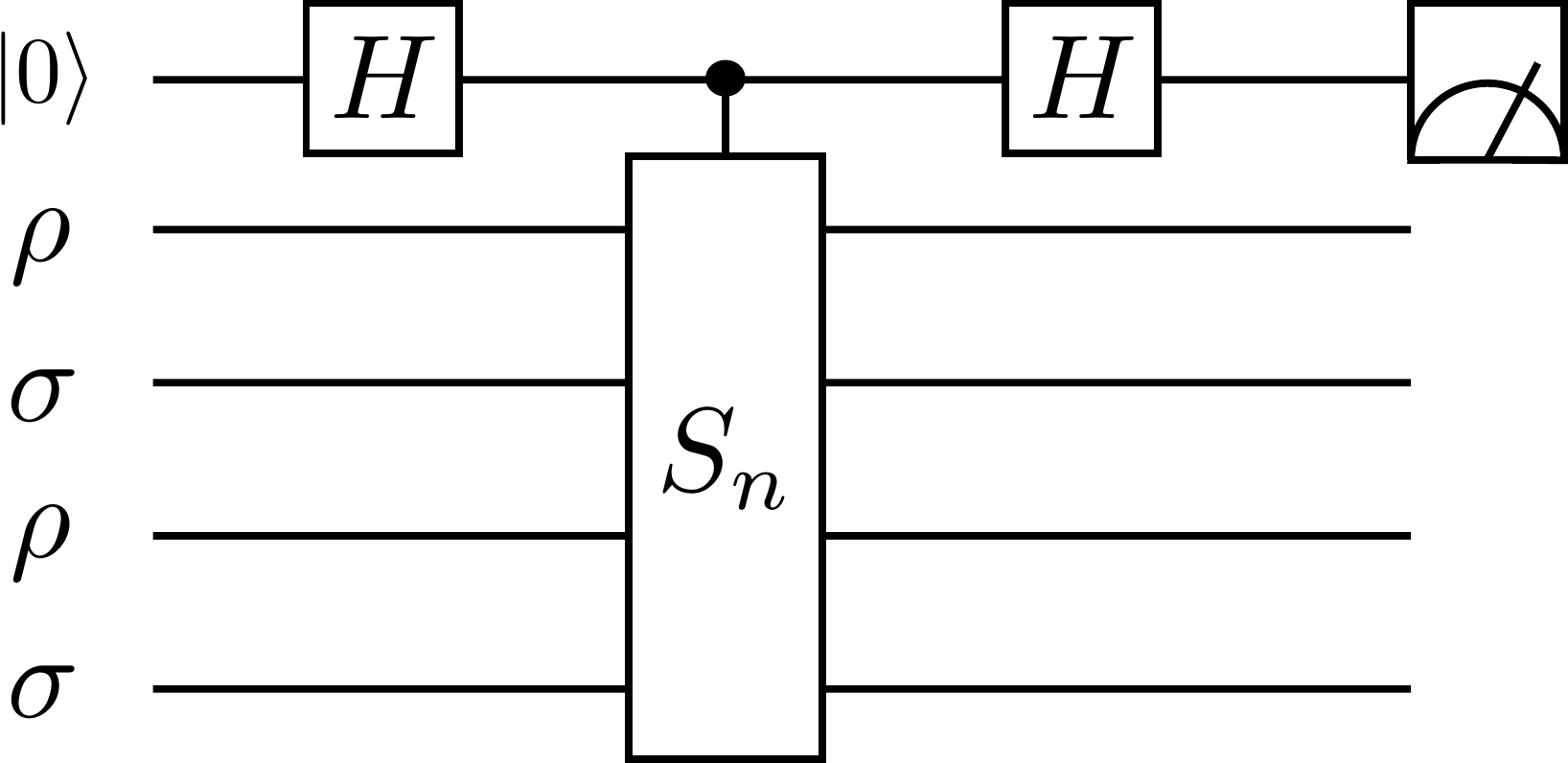}
    \caption{\textbf{The generalized SWAP test.} Using one ancillary qubit to control a cyclic permutation gate, one can estimate linear and non-linear functionals of a quantum state \cite{ekert2002direct,brun2004measuring}.}
    \label{fig:generalized-SWAP}
\end{figure}

Here, we show how to estimate functionals of the form $\Tr[\rho \sigma \rho \sigma]$, which are needed to compute the SSQFI bounds. The circuit in Fig. \ref{fig:generalized-SWAP} can be used to estimate this quantity for $n$-qubit quantum states $\rho$ and $\sigma$. In general, linear and non-linear functionals of quantum states can be directly estimated using what we refer to as the generalized SWAP test \cite{ekert2002direct,brun2004measuring}.

Let $\rho$ and $\sigma$ be $n$-qubit quantum states. Then in the standard orthonormal basis for $n$ qubits, we can write the first and second copies of the states as
\begin{align}
    \rho = \sum_{\vec{i}} a_{\vec{i}} \ket{\vec{i}}\bra{\vec{i}} , \quad \sigma = \sum_{\vec{j}} b_{\vec{j}} \ket{\vec{j}}\bra{\vec{j}}, \quad  \rho = \sum_{\vec{k}} c_{\vec{k}} \ket{\vec{k}}\bra{\vec{k}},\quad \sigma = \sum_{\vec{l}} d_{\vec{l}} \ket{\vec{l}}\bra{\vec{l}},   
\end{align}
where the boldface indices are bitstrings of the form $\vec{i}=i_1 i_2 \dotsm i_n$ with $i_1,i_2, \dots,i_n \in \{0,1\}$. Thus the initial state in Fig. \ref{fig:generalized-SWAP} is given as 
\begin{align}\label{eq:initialState}
    \rho_{\text{in}} = \sum_{\vec{i},\vec{j},\vec{k},\vec{l}} a_{\vec{i}}b_{\vec{j}}c_{\vec{k}}d_{\vec{l}} \ket{0,\vec{i},\vec{j},\vec{k},\vec{l}}\bra{0,\vec{i},\vec{j},\vec{k},\vec{l}}.
\end{align}
The Hadamard gate transforms the standard basis vectors as
\begin{align}
    H \ket{0} &= \frac{1}{\sqrt{2}}(\ket{0}+\ket{1}),\\
    H \ket{1} &= \frac{1}{\sqrt{2}}(\ket{0}-\ket{1}).
\end{align}
The cyclic shift operator (referred to as a permutation gate in the main text), $S_n$, is defined by its action on a tensor product basis for $n$-qubits:
\begin{align}
    S_n \ket{\psi_1,\psi_2,\dots,\psi_{n-1},\psi_n} = \ket{\psi_n,\psi_1,\psi_2,\dots,\psi_{n-1}}.
\end{align}
Note that in the case of $n=2$, the cyclic shift operator is simply the familiar SWAP operator \cite{nielsen_chuang}. Thus the action of the gate sequence in Fig. \ref{fig:generalized-SWAP} is:
\begin{align}
   & \quad \ket{0,\vec{i},\vec{j},\vec{k},\vec{l}}, \\
  \xrightarrow{\text{Hadamard}} & \quad \frac{1}{\sqrt{2}}(\ket{0,\vec{i},\vec{j},\vec{k},\vec{l}} +\ket{1,\vec{i},\vec{j},\vec{k},\vec{l}}), \\
  \xrightarrow{\text{controlled}-S_n} & \quad  \frac{1}{\sqrt{2}}(\ket{0,\vec{i},\vec{j},\vec{k},\vec{l}} +\ket{1,\vec{j},\vec{k},\vec{l},\vec{i},}),\\
  \xrightarrow{\text{Hadamard}} & \quad \frac{1}{2}(\ket{0,\vec{i},\vec{j},\vec{k},\vec{l}} +\ket{0,\vec{j},\vec{k},\vec{l},\vec{i},}).
\end{align}
Recalling Eq. \ref{eq:initialState}, one can see that the probability of measuring the ancilla qubit in the $\ket{0}$ state is 
\begin{align}
    p(0) &= \frac{1}{2}+\frac{1}{2}\sum_{\vec{i},\vec{j},\vec{k},\vec{l}} a_{\vec{i}}b_{\vec{j}}c_{\vec{k}}d_{\vec{l}} \braket{0,\vec{i},\vec{j},\vec{k},\vec{l}}{0,\vec{j},\vec{k},\vec{l},\vec{i}},\\
    &=\frac{1}{2}+\frac{1}{2}\sum_{\vec{i},\vec{j},\vec{k},\vec{l}} a_{\vec{i}}b_{\vec{j}}c_{\vec{k}}d_{\vec{l}} \delta_{\vec{ij}} \delta_{\vec{jk}} \delta_{\vec{kl}}\delta_{\vec{li}}, \\
    &= \frac{1}{2}+\frac{1}{2}\sum_{\vec{i}} a_{\vec{i}}b_{\vec{i}}c_{\vec{i}}d_{\vec{i}}, \\
    p(0) &= \frac{1}{2}+\frac{1}{2}\Tr[\rho \sigma \rho \sigma].
\end{align}
So, using a quantum computer with $4n+1$ qubits, one can directly estimate the functional $\Tr[\rho \sigma \rho \sigma]$ using the probability of measuring the ancillary qubit in the zero state. Moreover, removing the second copies of $\rho$ and $\sigma$, one is able to estimate functionals of the form $\Tr[\rho \sigma]$ using the same method. In that case, only $2n+1$ qubits are needed and the cyclic shift operator simply becomes the SWAP operator.

\section{Lower bound comparison}
\label{sec:comparison}
In this section, we first compare the conditions under which that the TQFI lower bound and the purity loss bound of Refs.~\cite{Modi16,Yang20} can saturate their inequalities and be equal to the QFI.  Then we present details on how the purity loss bound was computed in our heuristics.

\subsection{Bound saturation}

As shown below, while the TQFI bounds can always saturate the bounds (and be efficiently computable) for  states with low rank $r$, the purity loss bounds can never saturate the inequality for $r\geq 3$. This implies that the VQFIE bounds can always be tighter for low rank states with $r\geq 3$. 

First, let us recall that the TQFI bounds can be saturated if $m$ is equal to the rank $r$ of the probe state. Moreover, since one can estimate the truncated fidelities for $m\in\OC(\poly(n))$ this means that the TQFI bounds can be saturated for low rank states with $r\in\OC(\poly(n))$.

Let us now analyze the purity loss bound. In the main text we defined the quantity
\begin{align}
    \LC(\theta;\rho_{\theta})= 4\left(\Tr\left[\rho^2 G^2\right]-\Tr\left[\rho G \rho  G\right]\right)=2\sum_{i,j}(\lambda_i-\lambda_j)^2\abs{\bra{\lambda_i}G\ket{\lambda_j}}^2\,,
\label{eq:FragileStatebound}
\end{align}
which is a lower bound on the QFI as $\LC(\theta;\rho_{\theta})\leq I(\theta;\rho_{\theta})$. Then, from the definition of the quantum Fisher information~\cite{Hayashi_2004, Jing20}, we have
\begin{align}
    I(\theta;\rho_{\theta}) = 2\sum_{i,j}\frac{(\lambda_i-\lambda_j)^2}{\lambda_i+\lambda_j}\abs{\bra{\lambda_i}G\ket{\lambda_j}}^2\,.
\end{align}
From the fact that 
\begin{align}
    \frac{1}{\lambda_i+\lambda_j}\geq 1\,,
\end{align}
it is easy to see that $I(\theta;\rho_{\theta})$ takes its minimum when  
\begin{align}
    \lambda_i+\lambda_j=1\,.
\label{eq:lowerboundconditionNature}
\end{align}
The condition in~\eqref{eq:lowerboundconditionNature} implies that the QFI reaches its minima when $\rho_{\theta}$ is either a rank-$1$ or a rank-$2$ state. Hence, it is straightforward to see that if~\eqref{eq:lowerboundconditionNature} holds we have
\begin{align}
    I(\theta;\rho_{\theta})&=\LC(\theta;\rho_{\theta})\,.
\end{align}
However, for $r\geq 3$, the lower bound cannot saturate the inequality so that we have the following strict inequality
\begin{align}
    I(\theta;\rho_{\theta})>\LC(\theta;\rho_{\theta})~~\text{when}~~r\geq 3\,.
    \label{eq:NatureLowerboundRank}
\end{align}

\subsection{Heuristical computation of the purity loss}

In this section we describe additional details on how $\rho_{\text{ave}}$ was computed.  As shown in~\cite{Modi16} and as decribed in the main text, the lower bound in~\eqref{eq:FragileStatebound} can be approximated by  
\begin{align}\label{eq:PLapprox}
    \LC(\theta;\rho_{\theta})\approx 2\frac{\Delta\nu}{(\Delta x)^2}\,,
\end{align}
where
\begin{align}
    \Delta\nu=\Tr\left[\rho^2\right]-\Tr\left[\rho_{\text{ave}}^2\right]\,,
\label{eq:purityloss}
\end{align}
is the purity loss,  and where $(\Delta x)^2$ is the variance of the random variable $\theta(x)$, which defines the statistical fluctuation in the source.  In Ref.~\cite{Yang20}, the authors proposed computing the state $\rho_{\text{ave}}$ by using a stratified sampling technique. Namely, they assumed that $\theta(x)$ is drawn from a discretized Gaussian distribution $\GC$ with $K$ samples (or strata) and a variance $(\Delta x)^2$. That is,
\begin{equation}
    \rho_{\text{ave}}\approx \frac{1}{K}\sum_{j=1}^K \rho_{\theta_j}\,,
\end{equation}
where $\rho_{\theta_j}=W_{\theta_j}\rho W_{\theta_j}\ad$ and where $\theta_j$ is taken from $\GC$. Hence, the purity of $\rho_{\text{ave}}$  can be expressed as 
\begin{equation}
   \Tr[ \rho_{\text{ave}}^2]\approx \frac{1}{K^2}\sum_{j=1}^K \Tr[\rho_{\theta_j}^2]+\frac{2}{K^2}\sum_{j<m}^K \Tr[\rho_{\theta_j}\rho_{\theta_m}]\,,
\end{equation}
and can be efficiently computed via $(K^2+K)/2$ destructive swap tests~\cite{cincio2018learning} between the states $\rho_{\theta_j}$ and $\rho_{\theta_m}$ for $j\geq m$. 

As discussed in~\cite{Yang20}, the higher the number of strata $K$, the better the approximation  in~\eqref{eq:PLapprox}. Hence in order to determine how many strata we use in our numerics we here propose to determine $K$ so that the number of calls to a quantum computer is the same when computing the TQFI lower bound that when computing the purity-loss bound. 

First, let us determine how many calls to a quantum computer are necessary when computing the TQFI. As previously outlined, each iteration of VQFIE variationally diagonalizes the probe state and usually requires the use of a gradient descent algorithm over $p$ parameters in $V_{\vec{\beta}}$.  By employing the parameter shift-rule~\cite{mitarai2018quantum,schuld2019evaluating}, this requires to run $2p$ quantum circuits.  Moreover, the estimation of each gradient to a precision of $1/\sqrt{s}$ requires $s$ shots. In our numerical implementation, the training algorithm in VQSE took $t=200$. Hence, in each iteration of VQFIE where the sub-routine VQSE is ran we require $2stp$ calls to a quantum computer. Finally, we remark that to guarantee the trainability of VQSE we assume a hardware efficient ansatz with $\log(n)$ layers, meaning that $p=n\log(n)$. Then, as outlined in Ref. \cite{cerezo2019variational}, the fidelity computation requires one to estimate $m$ eigenvalues up to a precision of $1/\sqrt{s}$. A conservative estimate assumes $m=n$ eigenvalues are kept in the truncated state.  Finally, as described in the previous section, computing the TQFI bounds requires the estimation of the matrix elements of the $m$-by-$m$ symmetric matrix $T$ of Eq.~\eqref{eq:T-matrix} up to a precision $1/\sqrt{s}$. Here we recall that there are $(m^2+m)/2$ independent matrix elements in a symmetric matrix of size $m\times m$. Hence, in total, VQFIE requires 
\begin{align}
   \text{TQFI bound calls to quantum computer} = s  \left(2tn\log(n) + n + \frac{n+n^2}{2}\right)\,.
\end{align}

We turn now to the purity loss bound. This bound requires the estimation of  $\Tr[\rho^2]$ to a precision of $1/\sqrt{s}$, and also implies the computation of  $\Tr[\rho_{\text{avg}}^2]$. As previously discussed, computing the purity of $\rho_{\text{avg}}$ requires estimating $(K^2+K)/2$ state overlaps up to precision $1/\sqrt{s}$, where $K$ is the number of strata used for $\rho_{\text{avg}}$. Hence, the total number of calls to the quantum computer will be 
\begin{align}
   \text{Purity boundy calls to quantum computer} = s\left(\frac{K^2+K}{2}+1\right)\,.
\end{align}

Finally, the number of strata $K$ which leads to a fair comparison between the TQFI bound and the purity loss bound can be found by numerically solving 
\begin{equation}
    \frac{K^2+K}{2}+1= 2tn\log(n) + n + \frac{n+n^2}{2}\,.
\end{equation}

\end{document}